\documentclass[preprint,showpacs,preprintnumbers,amsmath,amssymb]{revtex4}

\usepackage{graphicx}
\usepackage{dcolumn}
\usepackage{bm}

\begin{document}

\title{Entropic sampling via Wang-Landau random walks in dominant energy subspaces}

\author{A. Malakis}
\altaffiliation[]{Corresponding author: amalakis@cc.uoa.gr}
\author{S. S. Martinos}
\author{I. A. Hadjiagapiou}
\author{N. G. Fytas}
\author{P. Kalozoumis}
\affiliation{Department of Physics, Section of Solid State
Physics, University of Athens, Panepistimiopolis, GR 15784
Zografos, Athens, Greece}

\date{\today}

\begin{abstract}
Dominant energy subspaces of statistical systems are defined with
the help of restrictive conditions on various characteristics of
the energy distribution, such as the probability density and the
forth order Binder's cumulant. Our analysis generalizes the ideas
of the critical minimum energy-subspace (CRMES) technique, applied
previously to study the specific heat's finite-size scaling. Here,
we illustrate alternatives that are useful for the analysis of
further finite-size anomalies and the behavior of the
corresponding dominant subspaces is presented for the 2D
Baxter-Wu, the 2D and 3D Ising models. In order to show that a
CRMES technique is adequate for the study of magnetic anomalies,
we study and test simple methods which provide the means for an
accurate determination of the energy - order-parameter ($E,M$)
histograms via Wang-Landau random walks. The 2D Ising model is
used as a test case and it is shown that high-level Wang-Landau
sampling schemes yield excellent estimates for all magnetic
properties. Our estimates compare very well with those of the
traditional Metropolis method. The relevant dominant energy
subspaces and dominant magnetization subspaces scale as expected
with exponents $\alpha/\nu$ and $\gamma/\nu$ respectively. Using
the Metropolis method we examine the time evolution of the
corresponding dominant magnetization subspaces and we uncover the
reasons behind the inadequacy of the Metropolis method to produce
a reliable estimation scheme for the tail-regime of the order
parameter distribution.
\end{abstract}

\pacs{05.50.+q, 64.60.Cn, 64.60.Fr, 75.10.Hk}

\maketitle

\section{Introduction}
\label{introduction}

Computer simulations based on Monte Carlo sampling methods have
increased dramatically our understanding of the behavior of the
standard classical statistical mechanics systems (for instance
Ising-like models), but also of the more complex systems, such us
disordered media, polymeric and glassy materials. Our main
approach, in the past half-century, was based on importance
sampling in the canonical ensemble. The Metropolis method and its
variants were, for many years, the main tools in condensed matter
physics, particularly for the study of critical
phenomena~\cite{metropolis53,bortz75,binder77,newman99,landau00}.
However, for complex systems effective potentials have a
complicated rugged landscape with many minima and maxima which
become more pronounced with increasing system size. In many such
cases the Metropolis method and its variants are very inefficient
methods. Entropic sampling methods (ESM) are alternatives to the
importance sampling methods, which at least in principle do not
suffer from such problems. Of course, for second order phase
transitions in unfrustrated systems cluster algorithms are quite
efficient and have essentially solved the ``critical slowing
down'' problem. The performance limitations of flat-histogram or
entropic methods have recently attracted considerable interest.
Even for simple systems, such as the Ising model, such methods
have tunnelling times in energy space that are not proportional to
$N^{2}$ as should be expected for a pure random walk but they are
proportional to a higher power. Moreover, it has been shown that
tunnelling times may be strongly depended on the model under
consideration~\cite{dayal04}. Furthermore, in order to apply an
ESM the density of states (DOS) of the system should be known.

Over the last decade, several efficient methods that directly
calculate the DOS of classical statistical models have been
developed. A few remarkable examples are the
entropic~\cite{landau00,lee93}, multicanonical \cite{berg92},
broad histogram~\cite{lima00,kastner00}, transition
matrix~\cite{wang02} and Wang-Landau~\cite{wang01} methods. Using
these methods, it is now possible to accurately estimate the DOS
of quite large classical statistical models~\cite{landau04}. Since
for complex systems the Metropolis method may be trapped for very
long times in non-representative energy subspaces it is reasonable
to consider an ESM program using an approximate DOS as an
alternative to the traditional importance sampling methods.
Although this idea exists in the literature~\cite{newman99}, the
effectiveness of the various possible implementations have not
been exploited by systematic comparative studies. For instance,
the following two-run strategy may be used in an ESM program. In
the first stage an accurate estimation of the DOS of the system
under consideration is achieved with the help of say the
Wang-Landau (WL) method, and in the second stage the derived DOS
is used in an ESM to estimate further properties of the system,
such as the order-parameter distribution. In such a two-stage
program, the CRMES method recently developed by Malakis \emph{et
al.}~\cite{malakis04,martinos05} may be used to restrict the
energy space and make such a program more efficient for the
estimation of the critical behavior of any statistical system. Our
first objective in this paper is to extent the CRMES condition and
to observe to what degree appropriately defined energy subspaces
are sufficiently large for the estimation of all finite-size
anomalies. Our second objective focuses on the possibility to
obtain all critical properties of the system by using a one-run
strategy of an ESM, based on a WL random walk in a restricted
energy subspace.

The rest of the paper is organized as follows. In
Sec.~\ref{section2} we provide a brief outline of the CRMES
restriction and give alternative definitions of the dominant
energy subspaces. The scaling of the extensions of these subspaces
is illustrated for the Baxter-Wu and Ising models. In
Sec.~\ref{section3a} we discuss how, by employing the WL and the
N-fold implementation of the WL scheme in a restricted energy
subspace, we may generate approximations of the DOS of the system
and at the same time obtain energy - order-parameter ($E,M$)
histograms. It is suggested that this one-run WL strategy yields
good estimates of all magnetic properties. In Sec.~\ref{section3b}
we consider the 2D Ising model as a test case. Our estimates are
compared with those of the traditional Metropolis method and the
expected magnetic scaling behavior is recovered. Finally, in
Sec.~\ref{section3c} we study our dominant subspace method in the
energy and the order-parameter space. The subspaces sufficient for
an accurate estimation of magnetic finite-size anomalies are
determined and their scaling is analyzed. The tail-regime of the
order-parameter critical distribution is briefly discussed  and
the shortcomings of the Metropolis method are clarified. Our
conclusions are summarized in Sec.~\ref{section4}.

\section{Alternatives for restricting the energy space}
\label{section2}

Let us start by recalling the original CRMES restriction of
Malakis \emph{et al.}~\cite{malakis04}. Assume that
$\widetilde{E}$ denotes the value of energy producing the maximum
term in the partition function of the statistical model, for
instance the Ising model, at some temperature of interest. Since
we deal with a finite system of linear size $L$, we are interested
in the properties (finite-size anomalies) near some pseudocritical
temperature $T_{L}^{\ast}$, which in general depends on $L$ but
also on the property studied. Thus, for the specific heat peaks
let us use the notation $T_{L}^{\ast}[C]\equiv T_{L,C}$ for the
pseudocritical temperature and define a set of approximations by
restricting the statistical sums to energy subranges around the
value
$\widetilde{E}=\widetilde{E}(T_{L}^{\ast}[C])=\widetilde{E}(T_{L,C})$.
Let these subranges of the total energy range $(E_{min},E_{max})$
be denoted as follows:
\begin{subequations}
\begin{equation}
\label{eq:1a} (\widetilde{E}_{-},\widetilde{E}_{+}),\;\;\;\;
\widetilde{E}_{\pm}=\widetilde{E}\pm \Delta^{\pm},\;\;\;\;\;
\Delta^{\pm}\geq0
\end{equation}
Accordingly, the peaks are approximated by:
\begin{equation}
\label{eq:1b} C_{L}(\widetilde{E}_{-},\widetilde{E}_{+})\equiv
C_{L}(\Delta^{\pm})=N^{-1}T^{-2}\left\{\widetilde{Z}^{-1}\sum_{\widetilde{E}_{-}}^{\widetilde{E}_{+}}E^{2}\exp{[\widetilde{\Phi}(E)]}\right.\\
\left.-
\left(\widetilde{Z}^{-1}\sum_{\widetilde{E}_{-}}^{\widetilde{E}_{+}}E\exp{[\widetilde{\Phi}(E)]}\right)^{2}\right\}
\end{equation}
\begin{equation}
\label{eq:1c} \widetilde{\Phi}(E)=[S(E)-\beta
E]-\left[S(\widetilde{E})-\beta
\widetilde{E}\right],\;\;\widetilde{Z}=\sum_{\widetilde{E}_{-}}^{\widetilde{E}_{+}}\exp{[\widetilde{\Phi}(E)]}
\end{equation}
\end{subequations}
Since by definition $\widetilde{\Phi}(E)$ is negative we can
easily see that for large lattices extreme values of energy (far
from $\widetilde{E}$) will have an extremely small contribution to
the statistical sums, since these terms decrease exponentially
fast with the distance from $\widetilde{E}$. It follows that, if
we request a specified accuracy, then we may restrict the
necessary energy range in which DOS should be sampled. To
introduce the minimum energy subspace (MES), we impose the
condition:
\begin{equation}
\label{eq:2}
\left|\frac{C_{L}^{\ast}(\Delta_{\pm})}{C_{L}^{\ast}}-1\right|\leq
r
\end{equation}
where $r$ measures the relative error and it will be set equal to
a small number ($r=10^{-6}$), and $C_{L}^{\ast}$ is the value of
the maximum of the specific heat obtained by using the total
energy range. With the help of a convenient definition, we can
specify the minimum energy subspaces satisfying the above
condition. Their finite-size extensions will be denoted by:
\begin{equation}
\label{eq:3} (\Delta\widetilde{E})_{C}\equiv
(\Delta\widetilde{E})_{C_{L}^{\ast},r}\equiv
min(\widetilde{E}_{+}-\widetilde{E}_{-})
\end{equation}
Demanding the same level of accuracy (r) for all lattice sizes, we
produce a size-dependence on all parameters of the above energy
ranges and in particular the extensions of the critical MES should
obey the scaling law~\cite{malakis04}:
\begin{equation}
\label{eq:4} \Psi_{C}\equiv
\frac{(\Delta\widetilde{E})^{2}_{C}}{L^{d}}\approx
L^{\frac{\alpha}{\nu}}
\end{equation}
In order to determine the location of the MES we may follow
successive minimal approximations to the specific heat peak
\cite{malakis04}:
\begin{subequations}
\label{eq:5}
\begin{equation}
\label{eq:5a} C_{L}(j)\equiv
C_{L}(\Delta_{j}^{-},\Delta_{j}^{+}),\;\;
\Delta^{\pm}_{j+1}=\Delta^{\pm}_{j}\pm\theta^{\pm}_{j+1},\;\;\
\Delta_{1}^{\pm}=0,\;\;\;\ j=1,2,...
\end{equation}
where one the above $\theta$-increments is chosen to be $1$ and
the other $0$ according to which side of $\widetilde{E}$ is
producing at the current stage the best approximation:
\begin{eqnarray}
\label{eq:5b}
&(\theta^{+}_{j+1}=1,\theta^{-}_{j+1}=0)\;\Leftrightarrow\;\mid
C_{L}-C_{L}(\Delta_{j}^{-},\Delta^{+}_{j}+1)\mid\leq\mid
C_{L}-C_{L}(\Delta_{j}^{-}+1,\Delta_{j}^{+})\mid\nonumber&\\
&(\theta^{+}_{j+1}=0,\theta^{-}_{j+1}=1)\;\Leftrightarrow\;\mid
C_{L}-C_{L}(\Delta_{j}^{-},\Delta^{+}_{j}+1)\mid>\mid
C_{L}-C_{L}(\Delta_{j}^{-}+1,\Delta_{j}^{+})\mid&
\end{eqnarray}
The above defines a sequence of relative errors for the specific
heat peaks ($r_{j}$):
\begin{equation}
\label{eq:5c} r_{j}=\left|\frac{C_{L}(j)}{C_{L}^{\ast}}-1\right|
\end{equation}
\end{subequations}
and the MES is the subspace centered at $\widetilde{E}$
corresponding to the first member of the above sequence
(\ref{eq:5}) satisfying: $r_{j}\leq r$. The location of these
subspaces can be predicted either by extrapolation, from smaller
lattices, or by using the early-stage DOS approximation of the WL
method~\cite{martinos05}. Using a sufficiently wider subspace we
can also accurately estimate their extensions and verify the
scaling law (\ref{eq:4}) as shown in Ref.~\cite{malakis04}.

The above scheme has been tested for the Ising and Baxter-Wu
models~\cite{malakis04,martinos05} and it has been shown that this
particular rule provides very accurate extensions satisfying quite
well the expected scaling law (\ref{eq:4}). However, one may
conceive other ways for specifying the locations of the energy
subspaces that will essentially produce comparable approximations.
A simple idea is to use a condition based on the energy
probability density ($f_{T_{L}^{\ast},C}(E)\propto
\widetilde{\Phi}(E)$) meaning the application of Eq.~(\ref{eq:6})
at a particular pseudocritical temperature $T_{L,C}$. That is, we
may define the end-points ($\widetilde{E}_{\pm}$) of the subspaces
by simply comparing the corresponding probability densities with
the maximum at the energy $\widetilde{E}$:
\begin{equation}
\label{eq:6} \widetilde{E}_{\pm}:\;\;\;
\exp{\{\widetilde{\Phi}(\widetilde{E}_{\pm})\}}\leq r
\end{equation}

Applying this restriction at a particular (pseudocritical)
temperature ($T_{L}^{\ast}$) is much easier than applying the
successive minimal approximations described in Eq.~(\ref{eq:5})
and it will produce comparable approximations for the specific
heat maxima. The corresponding scaling variable
$\Psi_{f_{T_{L}^{\ast},C}(E)}$ for the resulting MES should also
obey the law (\ref{eq:4}). Another interesting pseudocritical
temperature, is the temperature corresponding to the forth order
Binder's cumulant for the energy distribution. This reduced
cumulant is defined by:
\begin{equation}
\label{eq:7} V_{L}(T)=1-\frac{\langle E^{4}\rangle}{3\langle
E^{2}\rangle^{2}}
\end{equation}
and it is well known~\cite{martinos05,challa86,binder84} that this
quantity has a minimum, $V_{L}^{\ast}$, at a pseudocritical
temperature $T_{L}^{\ast}[V]\equiv T_{L,V}$. In the thermodynamic
limit this temperature also approaches the critical temperature
but for finite lattices is different from the pseudocritical
temperature $T_{L,C}$ of the specific heat. Thus, the maximum term
of the partition function, corresponding to the temperature of the
cumulant finite-size anomaly, will be now located at a different
value of the energy spectrum (say at $\widetilde{E}(T_{L,V})$).
Therefore, if we follow the probability density criterion
described above in Eq.~(\ref{eq:6}) to define the CRMES at these
temperatures we will generally find subspaces that do not coincide
with the subspaces for the specific heat. However, for large
lattices the non-overlapping parts of these subspaces are
relatively very small and we may run the WL algorithm in the union
of these subspaces in order to study both properties.

Let us now, follow an analogous approach with the successive
minimal approximations of Eq.~(\ref{eq:5}) for the cumulant
anomaly. It is of interest to note that if we define the CRMES by
a similar condition:
\begin{equation}
\label{eq:8}
\left|\frac{V_{L}^{\ast}(\Delta_{\pm})}{V_{L}^{\ast}}-1\right|\leq
r
\end{equation}
then, for sufficiently large lattices, the scaling law
(\ref{eq:4}) for the corresponding extensions (we may now use the
$\Psi_{V}=\frac{(\Delta\widetilde{E})^{2}_{V}}{L^{d}}$ variable)
will not be obeyed. This is due to the fact that the cumulant goes
asymptotically to a finite value
($V_{\infty}^{\ast}=V_{\infty}=2/3$) and therefore the relative
accuracy criterion (\ref{eq:8}) is not appropriate for large
lattices or, in other words, becomes ineffective (see also the
discussion bellow). We may now introduce the finite-size distance
from the asymptotic value of the Binder's parameter,
($V_{\infty}^{\ast}-V_{L}^{\ast}$), to our considerations and
modify the criterion (\ref{eq:8}) as follows:
\begin{equation}
\label{eq:9}
\left|\frac{V_{L}^{\ast}(\Delta_{\pm})}{V_{L}^{\ast}}-1\right|\left/
\left|\frac{V_{L}^{\ast}}{V_{\infty}^{\ast}}-1\right|\right.\leq r
\end{equation}

It appears (see bellow) that this option makes the extensions of
the resulting subspaces to follow very well the scaling law
(\ref{eq:4}) and for the 2D Ising model we obtain an almost
perfect coincidence with the extensions of the corresponding
subspaces obtained for the specific heat. The corresponding
scaling variable will be denoted by $\Psi_{V;V_{\infty}}=
\frac{(\Delta\widetilde{E})^{2}_{V;V_{\infty}}}{L^{d}}$. For a
second-order transition the limiting value of the energy cumulant
is known ($V_{\infty}^{\ast}=2/3$) and this makes the
implementation of the scheme (\ref{eq:9}) possible.

\begin{figure}
\includegraphics*[width=12 cm]{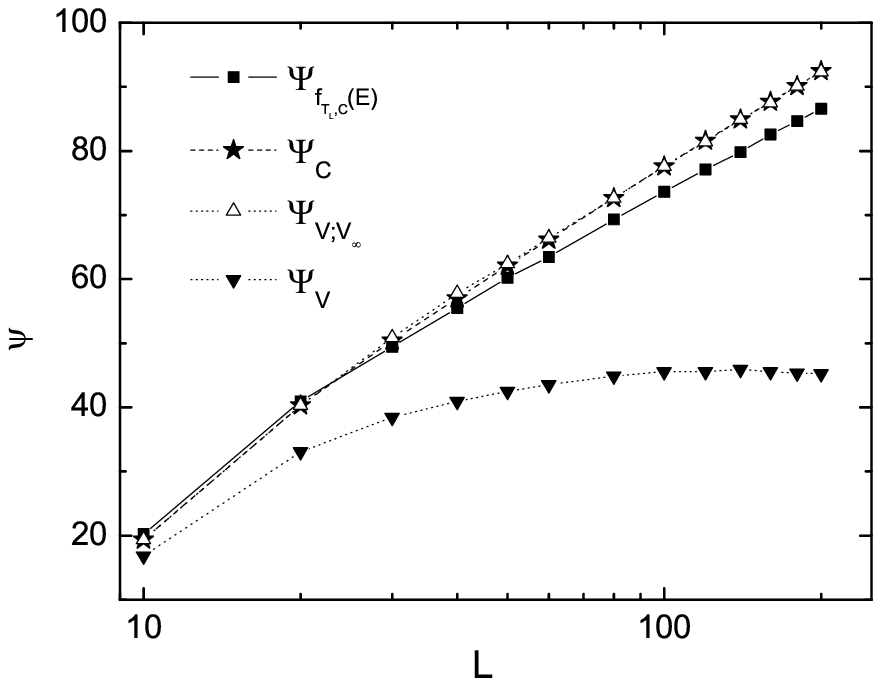}
\caption{\label{fig1}2D Ising model ($L=10-100$ \cite{malakis04},
$L=120,140,...,200$ new data). Behavior of scaling variables
corresponding to various alternative definitions of dominant
energy subspaces defined in the text.
$\Psi_{f_{T_{L}^{\ast},C}(E)}$ is the scaling variable defined
with the help of the restriction (\ref{eq:6}) at the
pseudocritical temperature $T_{L,C}$. Correspondingly, $\Psi_{C}$
is the scaling variable defined from Eq.~(\ref{eq:5}) at the
pseudocritical temperature $T_{L,C}$ and $\Psi_{V;V_{\infty}}$ is
the scaling variable defined from Eq.~(\ref{eq:9}) at the
pseudocritical temperature $T_{L,V}$. Finally, $\Psi_{V}$ is the
scaling variable defined from Eq.~(\ref{eq:8}) at the
pseudocritical temperature $T_{L,V}$. Note the strong decline of
this last variable from the expected asymptotic law, as explained
in the text. The log scale in x-axis is used in order to
facilitate the observation of the logarithmic behavior. }
\end{figure}

\begin{figure}
\includegraphics*[width=12 cm]{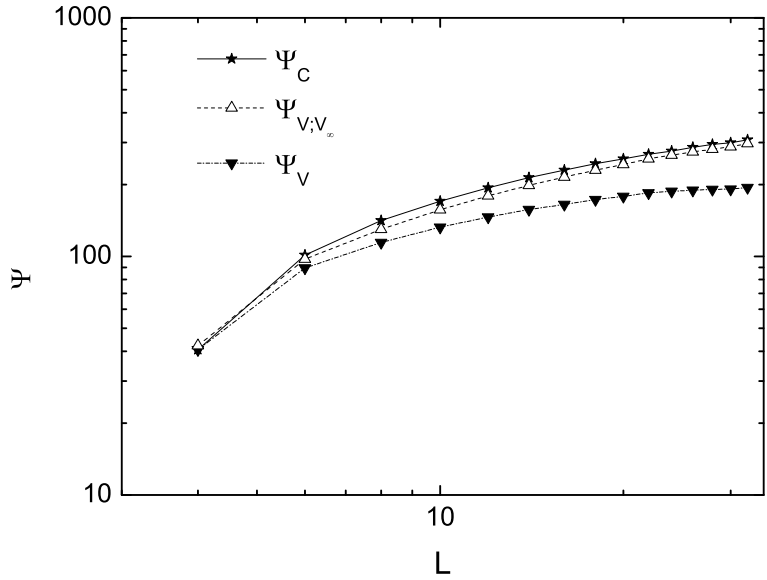}
\caption{\label{fig2}3D Ising model ($L=4-32$ \cite{malakis04}).
Behavior of the scaling variables $\Psi_{C}$,
$\Psi_{V;V_{\infty}}$ and $\Psi_{V}$, as in Fig.~{\ref{fig1}}.
Note again the clear decline of the variable $\Psi_{V}$, defined
by Eq.~(\ref{eq:8}). The log-log scale is used in order to observe
the expected power law.}
\end{figure}

\begin{figure}
\includegraphics*[width=12 cm]{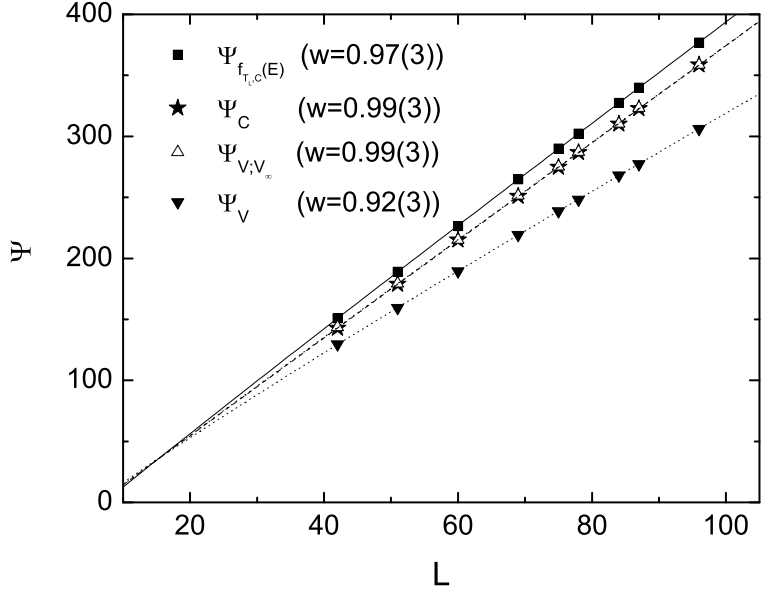}
\caption{\label{fig3}2D Baxter-Wu model
($L=42,51,60,69,75,78,84,87$ and $L=96$ \cite{martinos05}).
Behavior of the scaling variables $\Psi_{f_{T_{L}^{\ast},C}(E)}$,
$\Psi_{C}$, $\Psi_{V;V_{\infty}}$ and $\Psi_{V}$, again as in
Fig.~\ref{fig1}. The modest decline of the variable $\Psi_{V}$
defined in Eq.~(\ref{eq:8}) is due to the fact that the cumulant
minimum is much deeper for this model. The fitted lines correspond
to power laws of the form $\Psi=a+bL^{w}$ with exponents
$w=0.97(3)$, $w=0.99(3)$, $w=0.99(3)$ and $w=0.92(3)$,
respectively. The finest estimate to the expected critical
exponent $w=\alpha/\nu=1$ is the one corresponding to the original
definition (Eqs.~(\ref{eq:2})-(\ref{eq:5})) of the minimum
subspaces.}
\end{figure}

The above alternative definitions for the CRMES were applied to
the 2D Ising model using the DOS data of Ref.~\cite{malakis04}
$(L=10-100)$ but also new data up to $L=200\;
(L=120,140,160,180,200)$. Fig.~\ref{fig1} illustrates the scaling
behavior of the corresponding scaling variables for the CRMES of
the specific heat at its pseudocritical temperature $T_{L,C}$
using the minimum subspaces resulting from Eq.~(\ref{eq:5c}), and
also the minimum subspaces resulting, at this temperature, from
the probability density condition studied in Eq.~(\ref{eq:6}). The
same figure displays the estimates for the two options
(\ref{eq:8}) and (\ref{eq:9}) for the CRMES corresponding to the
minimum of the Binder's parameter at its
pseudocritical~temperature $T_{L,V}$. The expected logarithmic
scaling law~\cite{malakis04,ferdinand69} is obeyed well for all
definitions with a clear exception of the cumulant condition
(\ref{eq:8}) in which case the scaling variable shows a clear
decline from the logarithmic divergence for large lattices. This
is due to the fact that this parameter approaches the finite value
$V_{\infty}^{\ast}=2/3$, and the condition defined in
Eq.~(\ref{eq:8}) becomes ineffective for large lattices. Note that
with increasing lattice size several significant figures of this
finite number are determined from a small part of the dominant
energy subspace. The trick proposed in Eq.~(\ref{eq:9}) not only
keeps the scheme in the proper scaling form but also it is
remarkable that the resulting extensions of the CRMES of the
specific heat and the Binder's parameter completely coincide for
the 2D Ising model. This coincidence occurs also for the Baxter-Wu
model, but not for the 3D Ising model (see bellow). Note however,
that the corresponding pseudocritical temperatures are different
and their CRMES do not coincide (for $L=100$ their displacement is
$20$ energy levels), only their extensions are equal.
Fig.~\ref{fig2} illustrates the behavior of the same scaling
variables for the 3D Ising model using the DOS data of
Ref.~\cite{malakis04}. The situation is very similar and is
described by the power law (\ref{eq:4}) as discussed in
Ref.~\cite{malakis04}. Noteworthy, that the cumulant condition
(\ref{eq:8}) leads to a clear decline from the appropriate power
law for large lattices and the trick proposed in Eq.~(\ref{eq:9})
seems to yield the expected critical behavior. Finally,
Fig.~\ref{fig3} presents the analogous graphs for the Baxter-Wu
model using the DOS data of~\cite{martinos05}. Again the scaling
variables appear to follow the expected scaling law (\ref{eq:4})
and a decline in the case of the condition (\ref{eq:8}) is
observed. However, this decline is weaker for the Baxter-Wu model
due to the fact that the cumulant minimum is quite deeper for this
model, that is the difference $(V_{\infty}^{\ast}-V_{L}^{\ast})$
is relatively larger for the Baxter-Wu model. Fitting these data
to the expected power law (see the discussion in the caption of
Fig.~\ref{fig3}) we find that the best estimate for the exponent
$\alpha/\nu$ is produced from the
original~\cite{malakis04,martinos05} minimum subspaces of the
specific heat. As mentioned above the extensions of the CRMES
determined for the cumulant by the condition (\ref{eq:9}) for the
2D Baxter-Wu model coincide with the extensions of the CRMES of
the specific heat (Eq.~(\ref{eq:2})) of the model.

\section{Entropic Sampling via WL random walks}
\label{section3}
\subsection{Microcanonical estimators via the WL scheme}
\label{section3a}

As mentioned in the introduction, the ESM method using an exact
(if available) or an accurate approximation of the DOS may be
considered as an alternative to the various importance sampling
methods used in the literature to estimate canonical averages and
probability distributions of macroscopic thermodynamic variables.
Here we shall examine the idea of producing accurate estimates for
finite-size magnetic anomalies by using a simple ESM based on the
WL random walk in an appropriately restricted energy subspace
$(E_{1},E_{2})$. We shall also test our results by comparing to
the Metropolis method \cite{metropolis53}.

We implement a WL random walk in a restricted energy subspace
$(E_{1},E_{2})$ and at the same time we accumulate data for the
two-parameter ($E,M$) histogram. For a large system we employ a
multi-range algorithm~\cite{wang01} in which independent random
walks are used for different energy subintervals and the resultant
pieces are then combined to obtain the DOS in $(E_{1},E_{2})$. The
WL modification factor ($f_{j}$) is reduced at the jth iteration
according to: $f_{1}=e,\; f_{j}\rightarrow f_{j-1}^{1/2},\;
j=2,...,J_{fin}$ and for the histogram flatness we use a $5\%$
criterion as in our previous
studies~\cite{malakis04,martinos05,malakis04b}. Note that, the
detailed balance condition is satisfied in the limit $f\rightarrow
1$. Let the exact density of states be denoted by $G(E)$ and the
DOS of the above WL process be denoted by $G_{WL}(E)$. Similarly,
let $H_{WL}(E,M)$ be the histogram produced during the WL process
by a specific recipe which will be described below.

The resulting approximation of the DOS and the corresponding $E,M$
histograms may be used to estimate the magnetic properties of the
system in a temperature range, which is covered, by the restricted
energy subspace $(E_{1},E_{2})$. Canonical averages will be then
approximated as follows:
\begin{subequations}
\label{eq:10}
\begin{equation}
\label{eq:10a} \langle M^{n}\rangle=\frac{\sum_{E}\langle
M^{n}\rangle_{E}G(E)e^{-\beta E}}{\sum_{E}G(E)e^{-\beta E}}\cong
\frac{\sum_{E\in(E_{1},E_{2})}\langle
M^{n}\rangle_{E,WL}G_{WL}(E)e^{-\beta
E}}{\sum_{E\in(E_{1},E_{2})}G_{WL}(E)e^{-\beta E}}
\end{equation}
where the microcanonical averages $\langle M^{n}\rangle_{E}$ are
obtained from the $H_{WL}(E,M)$ histograms as:
\begin{equation}
\label{eq:10b}\langle M^{n}\rangle_{E}\cong\langle
M^{n}\rangle_{E,WL}\equiv
\sum_{M}M^{n}\frac{H_{WL}(E,M)}{H_{WL}(E)},\;\;
H_{WL}(E)=\sum_{M}H_{WL}(E,M)
\end{equation}
\end{subequations}
and the summation in $M$ runs over all values generated during the
process in the restricted energy subspace $(E_{1},E_{2})$.

The accuracy of the magnetic properties obtained from the above
averaging process will depend on many factors. Firstly, the used
energy subspace restricts the temperature range for which such
approximations may be accurate. This restriction has as a result
that the process will not visit all possible values of $M$, but
this fact is of no consequence for the accuracy of the magnetic
properties at the temperature range of interest, as far as the
estimated DOS is accurate. Secondly, the accuracy of the above
microcanonical estimators will, as usually, depend on the total
number of visits to a given energy level $(H_{WL}(E))$, and also
to the number of different spin states visited within this energy
level. However, these are statistical fluctuations inherent in any
Monte Carlo method and we should expect improvement by increasing
the number of repetitions of the process. At this point let us
point out that statistical fluctuations may be reduced, as
usually, by multiple measurements but also by using some
refinements of the original WL algorithm~\cite{zhou05,lee}. An
illustrative figure including such a refinement will be presented
in the next subsection. Finally, the construction of reliable
(uniform) approximations for microcanonical averages is an
important open problem, discussed recently in some detail by P. M.
C. de Oliveira~\cite{oliveira00}. The microcanonical
approximations $(\langle M^{n}\rangle_{E,WL})$, appearing in
Eq.~(\ref{eq:10}), and used in this paper are obtained from the WL
multi-range process, and the corresponding N-fold version of Shulz
\emph{et al.}~\cite{schulz01}. The recipes employed are outlined
and tested in the sequel.

There are mainly three categories of microcanonical simulation
approaches. In the first, one tries to satisfy completely the
fixed-energy constraint, a typical example is the Q2R cellular
automaton~\cite{pomeau84,herrmann86}. In the second, one tries to
mildly relax the energy constraint by using relatively small
energy-windows in order to avoid ergodic problems, as done by
Creutz~\cite{creutz83} in his ``demon'' method. Finally, the
fixed-energy constraint is completely relaxed \cite{berg93} and
various ideas have been tried, ranging from the tuned temperature
canonical approach of Oliveira~\cite{oliveira00} to the flat
histogram recipe of J.-S. Wang~\cite{jswang99}.

In order to present some justification for the scheme of
Eqs.~(\ref{eq:10a})-(\ref{eq:10b}) let us suppose that we know the
exact DOS and we are about to perform an infinitely long entropic
sampling in the restricted energy subspace $(E_{1},E_{2})$. In
this case we can write for the Monte Carlo process with sampling
probability $P_{i}\propto 1/G(E)$~\cite{newman99}:
\begin{subequations}
\label{eq:11}
\begin{equation}
\label{eq:11a} \langle
M^{n}\rangle=\frac{\sum_{(E_{1},E_{2})}\langle
M^{n}\rangle_{E}G(E)e^{-\beta
E}}{\sum_{(E_{1},E_{2})}G(E)e^{-\beta E}}\cong
\frac{\sum_{i=1}^{\mathcal{N}}M_{i}^{n}P_{i}^{-1}e^{-\beta
E}}{\sum_{i=1}^{\mathcal{N}}P_{i}^{-1}e^{-\beta E}}\cong
\frac{\sum_{(E_{1},E_{2})}\langle
M^{n}\rangle_{E,ESM}G(E)e^{-\beta
E}}{\sum_{(E_{1},E_{2})}G(E)e^{-\beta E}}
\end{equation}

The last approximation in Eq.~(\ref{eq:11a}) assumes that in the
limit of an infinite Markovian chain the $H_{ESM}(E)$ histogram is
perfectly flat. Accordingly, $H_{ESM}(E)$ has been set equal to a
constant in the denominator (in replacing the sum over the
$\mathcal{N}$ sampled microstates by a sum over energies) and then
moved to the numerator in order to form the ESM microcanonical
averages defined by:
\begin{equation} \label{eq:11b}
\langle M^{n}\rangle_{E,ESM}\equiv
\sum_{M}M^{n}\frac{H_{ESM}(E,M)}{H_{ESM}(E)},\;\;
H_{ESM}(E)=\sum_{M}H_{ESM}(E,M)
\end{equation}
\end{subequations}
The above observation shows that the ESM microcanonical average
should be an unbiased estimator for the fundamental microcanonical
average:
\begin{equation} \label{eq:12}
\langle M^{n}\rangle_{E,ESM}\stackrel{\mathcal{N}\rightarrow
\infty}{\longrightarrow}\langle M^{n}\rangle_{E}
\end{equation}

Therefore, Eq.~(\ref{eq:12}) provides the essential theoretical
support for using a multi-range WL process (at its late stages) to
obtain microcanonical simulators. It is reasonable to expect that
the high-level stages $(j\gg 1)$ of the WL process will resemble
the dynamics of the ESM and therefore will produce good
approximations for the microcanonical averages. This approach is
similar in many respects to the flat histogram method of J.-S.
Wang~\cite{jswang99}. Since the resemblance of the WL process with
the ESM depends on the value of the control parameter $(f_{j})$,
we shall classify our recipes by using the $j$-range utilized for
updating the $(E,M)$ histogram during the WL process. Thus, if all
accepted microstates of the WL process during the $j$-range
$(j=J_{init},...,J_{fin})$ are used, for updating the $(E,M)$
histogram, the resulting recipes will be denoted by
WL$(J_{init}-J_{fin})$. When using the N-fold version of the WL
process we always start with several $(J_{WL})$ WL $j$-iterations
and then continue the process from the next level
$(J_{N-fold}=J_{WL}+1)$ by carrying out further N-fold WL
iterations $(j=J_{N-fold},...,J_{fin})$. In this recipe, we shall
use only the N-fold iterations for updating the $(E,M)$ histogram
and time-weight the histogram by using the life-time of
microstates calculated according to the N-fold
method~\cite{bortz75,schulz01,malakis04b}. The resulting recipes
are denoted by WL(N-fold)$(J_{N-fold}-J_{fin})$.

The above described schemes were tested by using Eq.~(\ref{eq:10})
to obtain certain magnetic properties of the 2D Ising model (for
lattices with sizes $L=20-100$) and compare to the estimates
obtained by the Metropolis method. The finite-size anomaly of the
susceptibility and its value at the exact critical temperature, as
well as the value of the order parameter also at the exact
critical temperature were used in these tests and appear in the
following subsection. For the magnetic susceptibility we have used
the definition:
\begin{equation}
\label{eq:13} \chi_{L}(T)=\frac{1}{N}\left(\frac{\langle
M^{2}\rangle-\langle |M|\rangle^{2}}{T}\right)
\end{equation}
and for the order parameter:
\begin{equation}
\label{eq:14} m_{T}\equiv \langle |M|/N\rangle_{T}
\end{equation}

\subsection{Metropolis versus WL - CRMES schemes. A comparative study}
\label{section3b}

The estimates of the Metropolis method were obtained as follows.
First an initial equilibration period of $50\times L^{2}$ usual
Monte Carlo steps (lattice sweeps) was applied without updating
the histograms. After thermalization, the updating of the
histograms was applied in every Monte Carlo step, while the
magnetic properties, the order-parameter distributions and the
time evolving dominant $M$-subspaces were determined and observed
in time steps of $50\times L^{2}$ Monte Carlo sweeps. The time
evolution lasted a total of $300$ such time steps for all lattice
sizes (for $L=70, 120$ and $L=140$, see also the discussion
bellow). Thus, for a lattice of linear size $L=100$, the above
time evolution accounts to a total of $1.5\times 10^{8}$ lattice
sweeps.

The estimates of the WL multi-range process were obtained using,
for each lattice size, $30$ independent random walks in the
appropriate ($E_{1},E_{2}$) energy subspace. These subspaces were
chosen carefully to cover a wide temperature range close to the
critical point, so that the DOS and the $H(E,M)$ histograms
produced would yield accurate estimates of all thermal and
magnetic properties in the temperature-range. The energy subspaces
were wide enough to produce relative accuracies , within the
scheme, far beyond the criterion $r=10^{-6}$, which was finally
applied to determine the dominant subspaces. The time requirements
of the described Metropolis estimation were notably greater than
the described WL multi-range process of $30$ independent random
walks. For $L=100$ the WL scheme was about $3$ times faster than
the Metropolis scheme. Note that the Metropolis method produces
estimates only for one particular temperature and not for a wide
temperature-range. For a lattice of linear size $L=140$ the above
described WL scheme is at least $10$ times faster than the
described Metropolis scheme.

\begin{figure}
\includegraphics*[width=12 cm]{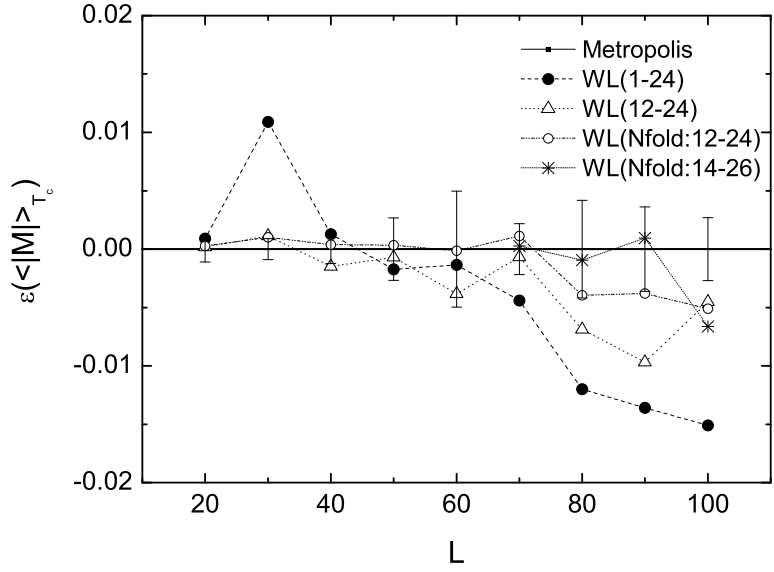}
\caption{\label{fig4} Relative deviations of WL ES schemes, with
respect to the Metropolis algorithm defined in Eq.~(\ref{eq:15}),
calculated from the order parameter at the critical temperature
$T_{c}$. The error bars used show the Metropolis relative
uncertainties calculated as $5$ standard deviations in the
equilibrium regime.}
\end{figure}

\begin{figure}
\includegraphics*[width=12 cm]{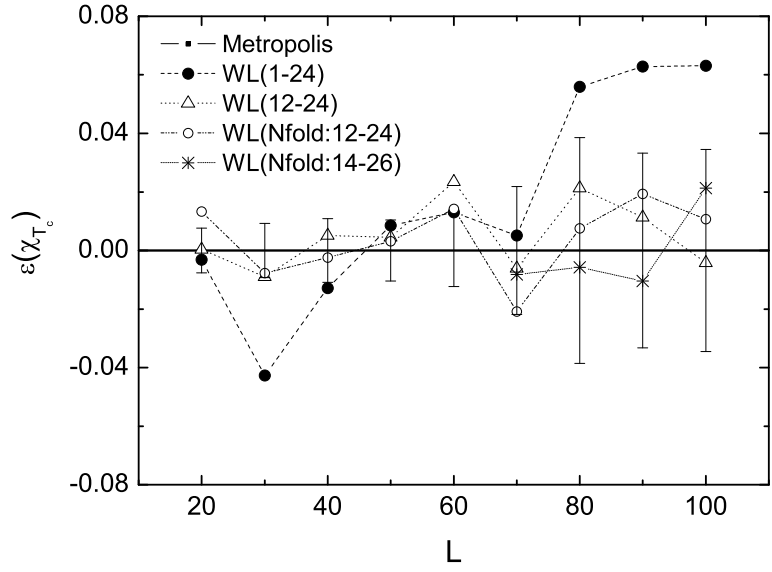}
\caption{\label{fig5} The same deviations, as in Fig.~\ref{fig4},
again at $T_{c}$, calculated now from the susceptibility. The
error bars as in Fig.~\ref{fig4}, show again the Metropolis
relative uncertainties calculated as $5$ standard deviations in
the equilibrium regime.}
\end{figure}

\begin{figure}
\includegraphics*[width=12 cm]{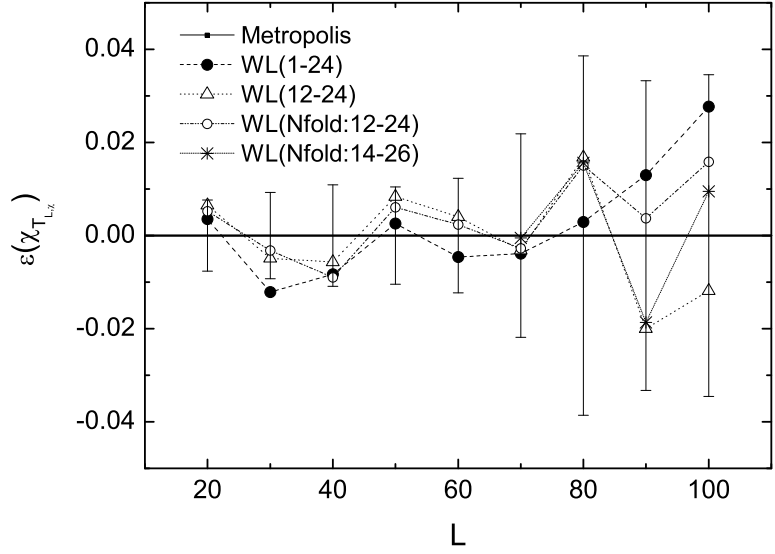}
\caption{\label{fig6} Susceptibility deviations at the
corresponding pseudocritical temperatures. The error bars as in
Figs.~\ref{fig4},\ref{fig5}.}
\end{figure}

The estimates derived from the WL recipes appear to have
relatively small deviations from the corresponding Metropolis
estimates and, with a notable exception, they seem to be within
the given error bounds. The comparison with the Metropolis method
is presented in Figs.~\ref{fig4},\ref{fig5} and \ref{fig6}. The
relative variations shown in these graphs are defined with respect
to the Metropolis estimates by assuming that these are more
accurate, i.e we define:
\begin{equation}
\label{eq:15} \varepsilon(Q)=\frac{Q_{Metr}-Q_{WL}}{Q_{Metr}}
\end{equation}
where $Q=m_{T}$ and/or $Q=\chi_{T}$. The error bounds used in
these graphs refer to the Metropolis estimates in the observed
equilibrium regime which is roughly defined as the last almost
flat part ($t\cong150-300$) in the above described $300$
time-steps. To define the error bounds we have used a confidence
level of $5$ standard deviations obtained in this wide
time-window.

Fig.~\ref{fig4} shows the relative deviations from the Metropolis
mean values of the order-parameter at the exact critical
temperature for four recipes of the WL scheme, as indicated in
this figure. The Metropolis estimates have very small error bounds
of the order of $0.5\%$ or less, besides the fact that we have
applied the above mentioned demanding confidence limit. The
deviations of the WL schemes are reasonable (of the order of $1\%$
or less) with the clear exception of the case WL(1-24) in which
the histogram's $(H(E,M))$ updating started from the very early
stage of the WL process. For large lattices, this recipe seems to
produce a significant overestimation of the order-parameter, at
the exact critical temperature, and this is enhanced with
increasing lattice size. Since, the detailed balance is strongly
violated at the early stages of the WL process, the observed
failure of this recipe should be attributed to the detailed
balance violation. The related overestimation may be possibly a
result of an oversampling distortion of large magnetization values
at the low energy part of the energy range used. Distortions
stemming from the violation of the detailed balance condition are
difficult to explain and in general their origin is a subtle
matter. However, such systematic distortions are not observed for
the other three recipes appearing in Fig.~\ref{fig4}. We consider
this as a first strong indication that the weak violation of the
detailed balance condition for these high-level WL schemes is not
statistically significant. The behavior of the relative deviations
for the susceptibilities at the exact critical temperature in
Fig.~\ref{fig5}, and at the pseudocritical temperatures in
Fig.~\ref{fig6} appear in general, much better from those shown in
Fig.~\ref{fig4}. Again the distortions of the WL(1-24) scheme
become pronounced as we increase the lattice size.

\begin{figure}
\includegraphics*[width=12 cm]{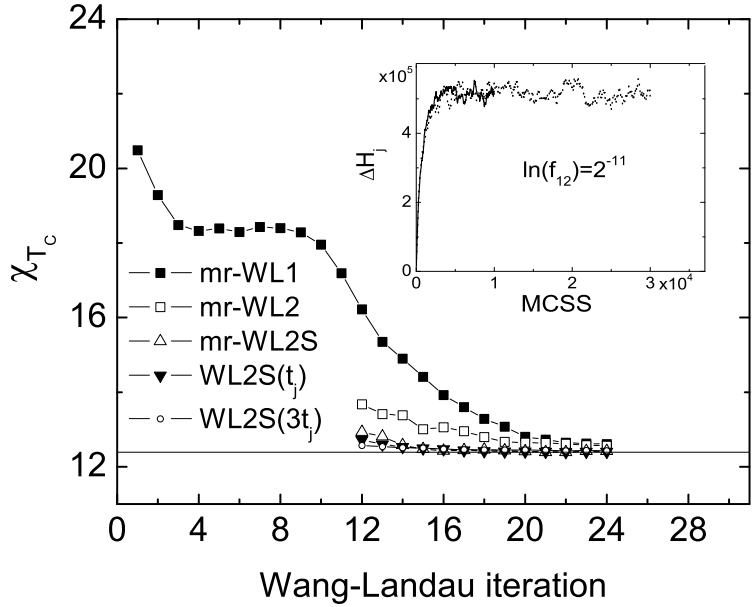}
\caption{\label{fig7} Evolution of the estimates of the critical
susceptibility with the WL iteration on a lattice of linear size
$L=30$. The filled (open) squares present the case mr-WL1 (mr-WL2)
which is the multi-range WL(1-24) (WL(12-24)) recipe, described in
the text. Upper open triangles (mr-WL2$\mathcal{S}$) illustrate
the evolution when a separation ($\mathcal{S}$) refinement of 16
spin-flips is applied between successive records in the ($E,M$)
histograms. The last two cases, WL2$\mathcal{S}(t_{j}$) and
WL2$\mathcal{S}(3t_{j}$) (down filled triangles and open circles
respectively), correspond to a simple one-range approach of
different simulation times. The duration of each WL iteration was
carefully chosen, so that saturation of the histogram fluctuation
was well obeyed, as shown in the inset ($j=12$). The solid line
represents the estimates of $\chi_{T_{c}}$ obtained by the
Metropolis run, as explained in the text.}
\end{figure}

At this point let us try to observe in more detail the effect of
the WL iteration level on an estimated magnetic property and also
the effect of one of the simplest refinements of the WL algorithm
for the square Ising lattice of size $L=30$. Using an accurate
DOS, $G_{WL}$, obtained from a previous WL run ($J_{fin}=24$), we
have calculated, with the help of Eq.~(\ref{eq:11}), the critical
susceptibility $\chi_{T_{c}}$ as a function of the WL iteration
level in a new WL diffusion process. In this new multi-range (mr)
process each WL iteration level is repeated $30$ times for each
energy subinterval. Fig.~\ref{fig7} shows the evolution of the
critical susceptibility to its equilibrium value for five cases.
In the first case (mr-WL1) the updating of $(E,M)$ histograms
follows the recipe WL($1-24$) and in the second case (mr-WL2) the
recipe WL($12-24$). The third case (mr-WL2$\mathcal{S}$) follows a
refinement of the WL algorithm proposed by Zhou and
Bhatt~\cite{zhou05}. The additional element of the algorithm is
that now we allow for a number $\mathcal{S}$ ($\mathcal{S}=16$)
spin-flips between successive records in the histograms (and in
the DOS modification). Introduction of such a separation
diminishes systematic errors due to the correlation between
adjacent records as shown in Ref.~\cite{zhou05}. From the first
case we note that starting the $(E,M)$ histogram updating process
at the early stage of the WL difusion is analogous to adding a
``non-equilibrium memory effect'' in our cumulative histograms.
This early effect is stronger when the WL algorithm is used in a
simple one-range fashion, as our simulations have shown. It is
also apparent that the refinement introduced by the separation
$\mathcal{S}$ clearly improves, in the cost of the additional
spin-flips, the behavior of the evolution of the magnetic
susceptibility $\chi_{T_{c}}$ towards its final equilibrium value,
which otherwise ($\mathcal{S}=1$) is obtained in a longer run.

Zhou and Bhatt~\cite{zhou05} have given a proof of the convergence
of the WL algorithm and found that the fluctuation of the
histogram is proportional to $1/\sqrt{\ln{f}}$ where $f$ is the
modification factor. This has been recently confirmed by numerical
tests~\cite{lee}, where it was shown that the criterion for
reducing $f$ goes beyond the ``flat histogram'' idea and that
``enough statistics'' should be obtained in each WL iteration.
According to Lee, Okabe, and Landau~\cite{lee} an optimal
algorithm will stop the simulation as soon as the histogram
fluctuation at the jth iteration, denoted as $\Delta
H_{j}=\sum_{E} (H_{j}(E)-\min_{E}\{H_{j}(E)\})$, becomes
saturated. In order to observe whether we will have a noticeable
improvement on the magnetic susceptibility estimates by applying
the proposed entropic sampling scheme (WL(12-24)) for longer
simulation times, at each $j(=12-24)$ iteration, we present also
in Fig.~\ref{fig7} the cases WL2$\mathcal{S}(t_{j}$) and
WL2$\mathcal{S}(3t_{j}$). Both these runs were performed via a
simple one-range WL scheme, using at each iteration more Monte
Carlo steps per spin (MCSS) than required for the saturation of
$\Delta H_{j}$. Moreover, the latter case uses three times the
number of MCSS of the former. The inset of Fig.~\ref{fig7}
illustrates the clear saturation of $\Delta H_{j=12}$ for these
runs (the solid (dotted) line corresponds to duration $t_{j}$
($3t_{j}$)). Comparing the estimates of the last three cases shown
in Fig.~\ref{fig7} ($\mathcal{S}=16$) we may draw the following
conclusions. Firstly, the multi-range approach applied in the case
mr-WL2S gives comparable estimates to those obtained from the
one-range implementation of the WL scheme and secondly, a much
longer run, such as the WL2$\mathcal{S}(3t_{j}$), does not yield a
noticeable improvement. It appears that an optimal and quite
accurate implementation of the proposed entropic scheme can be
constructed using a multi-range $H(E,M)$ histogram updating
process, during the high-level ``well saturated'' WL iterations.

The recent combination of Lee's entropic sampling with the WL
algorithm presented in Ref.~\cite{lee} may be also implemented to
test and possibly improve the proposed here CRMES entropic scheme.
Finally, the adaptive algorithm of Trebst \textit{et
al.}~\cite{trebst04} is of particular interest for a CRMES
implementation and would be also stimulating to compare the
``bottleneck region'', or region of minimum diffusivity, of this
method with the dominant energy subspaces as defined in this
manuscript. Therefore, we conclude that, the high-level WL(CRMES)
schemes are reasonable alternatives to the Metropolis method. The
estimates for thermodynamic parameters involving higher moments of
critical distributions appear to be of the same or even better
accuracy with those corresponding to the traditional method.

\subsection{Energy and order-parameter dominant subspaces}
\label{section3c}

The energy $\widetilde{E}$ producing the maximum term in the
partition function at the pseudocritical temperature of the
susceptibility $T_{L}^{\ast}[\chi]\equiv T_{L,\chi}$ may be easily
located. Thus, we may apply a minimal approximation scheme
analogous to that of Eq.~(\ref{eq:5}) to observe the evolution of
the susceptibility maximum as we expand the energy subspaces
centered around $\widetilde{E}$. Now the value of the
susceptibility is used in place of the specific heat and the
resulting CRMES is the subspace centered at $\widetilde{E}$
corresponding to the first member of the sequence satisfying:
\begin{equation}
\label{eq:16}\left|\frac{\chi_{L}^{\ast}(\Delta_{\pm})}{\chi_{L}^{\ast}}-1\right|\leq
r
\end{equation}
Provided that our initial guess for $(E_{1},E_{2})$ is wide enough
we also obtain accurate estimates for the finite-size extension of
these subspaces and we would expect that the relevant scaling
variable $\Psi_{\chi}=\frac{(\Delta
\widetilde{E})^{2}_{\chi}}{L^{d}}$ would obey the scaling law
(\ref{eq:4}). Therefore, we may restate the scaling law as:
\begin{equation}
\label{eq:17} \Psi_{\Theta}\equiv
\frac{(\Delta\widetilde{E})^{2}_{\Theta}}{L^{d}}\approx
L^{\frac{\alpha}{\nu}}
\end{equation}
where $\Theta(\equiv\Theta_{L})$ denotes the finite-size value of
some thermodynamic variable close to a critical point and
$(\Delta\widetilde{E})_{\Theta}$ is the extension for
appropriately defined minimum energy subspaces. In analogy with
our findings~\cite{malakis04,martinos05} for a diverging specific
heat behavior, we expect that a diverging susceptibility and the
criterion (\ref{eq:16}) will yield extensions scaling according to
the above law. The alternative method described in
Sec.~\ref{section2} which employs the energy density function (see
Eq.~\ref{eq:6}) may be also used and is easier to implement.

\begin{figure}
\includegraphics*[width=12 cm]{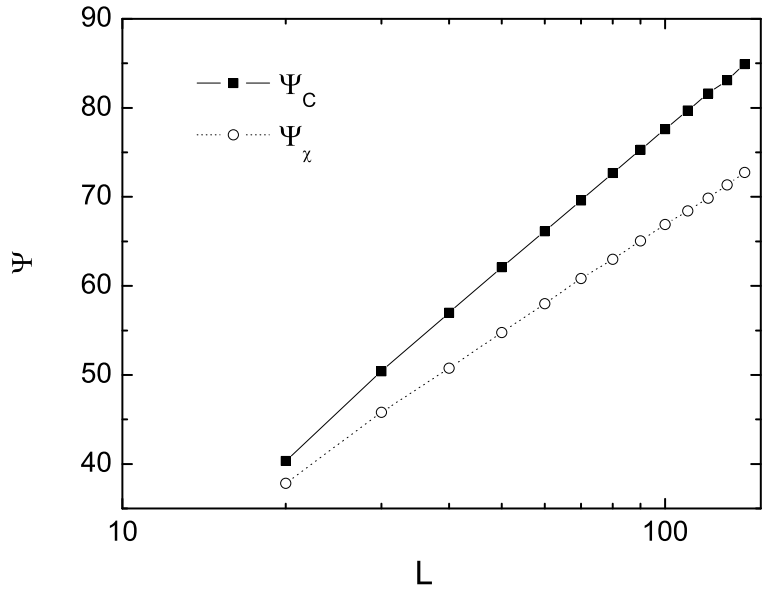}
\caption{\label{fig8} 2D Ising model: Logarithmic scaling behavior
of the finite-size extensions of the CRMES defined with the help
of specific heat maxima according to the successive minimal
approximations (Eq.~(\ref{eq:5})) and the analogous extensions of
the CRMES defined with the help of the susceptibility maxima
(Eq.~(\ref{eq:16})).}
\end{figure}

Fig.~\ref{fig8} illustrates  the scaling behavior of the extension
of the CRMES defined with the help of the specific heat maxima
(Eq.~(\ref{eq:5})) and the corresponding CRMES defined with the
help of the magnetic susceptibility maxima, as discussed above
(Eq.~\ref{eq:16}). The corresponding scaling variables should be
expected to obey the scaling law (\ref{eq:17}), and for the 2D
Ising model the well known logarithmic
law~\cite{malakis04,ferdinand69}. As seen from this figure, this
logarithmic law is well obeyed for both restrictive schemes as
expected.

Finally, we may describe a procedure for specifying the dominant
subspace in the order-parameter space. We assume that the energy
subspace $(E_{1},E_{2})$ is sufficiently broad to approximate to
the desired degree of accuracy the probability density of the
order-parameter at some temperature of interest by:
\begin{equation}
\label{eq:18}
P_{T}(M)\cong\frac{\sum_{E\in(E_{1},E_{2})}\frac{H_{WL}(E,M)}{H_{WL}(E)}G_{WL}(E)e^{-\beta
E}}{\sum_{E\in(E_{1},E_{2})}G_{WL}(E)e^{-\beta E}}
\end{equation}
Next, we find the value $\widetilde{M}$ that maximizes the above
density, at the pseudocritical temperature $T_{L,\chi}$ (or some
other temperature, for instance the exact critical temperature),
and we locate the end-points $(\widetilde{M}_{\pm})$ of the
magnetic critical subspaces by:
\begin{equation}
\label{eq:19} \widetilde{M}_{\pm}:\;\;\;
\frac{P_{T_{L,\chi}}(\widetilde{M}_{\pm})}{P_{T_{L,\chi}}(\widetilde{M})}\leq
r
\end{equation}
The above condition is in full analogy with that of
Eq.~(\ref{eq:6}) applied there in the energy space and since we
will now consider only dominant $M$-subspaces, defined with the
help of the above probability density criterion, we shall avoid in
our notation the explicit reference to the probability density.
Accordingly, we denote the extension of the resulting magnetic
subspaces by:
\begin{equation}
\label{eq:20} (\Delta \widetilde{M})_{T_{L,\chi}}\equiv(\Delta
\widetilde{M})_{P_{T_{L,\chi}}(M)}\equiv
\min{(\widetilde{M}_{+}-\widetilde{M}_{-})}
\end{equation}
and  we should now examine for a scaling law of the form:
\begin{equation}
\label{eq:21}
\Xi_{T_{L,\chi}}\equiv\Xi_{P_{T_{L,\chi}}(M)}\equiv\frac{(\Delta\widetilde{M})^{2}_{T_{L,\chi}}}{L^{d}}\approx
L^{\frac{\gamma}{\nu}}
\end{equation}

\begin{table*}
\caption{\label{table1}$\Xi=\frac{(\Delta M)^2}{L^2}\simeq
aL^{w}$, exact $w=\frac{\gamma}{\nu}=1.75$. Data fitted:
$L=50-100$, $T=T_{c}$.}
\begin{ruledtabular}
\begin{tabular}{lcccc}
  &Metropolis &WL(1-24) &WL(12-24) &WL(N-fold:14-26) \\
\hline a &0.59(6) &0.31(9) &0.54(1) &0.54(1)\\  w
&1.73(2) &1.89(6) &1.76(1) &1.76(1)\\
\end{tabular}
\end{ruledtabular}
\end{table*}

\begin{figure}
\includegraphics*[width=12 cm]{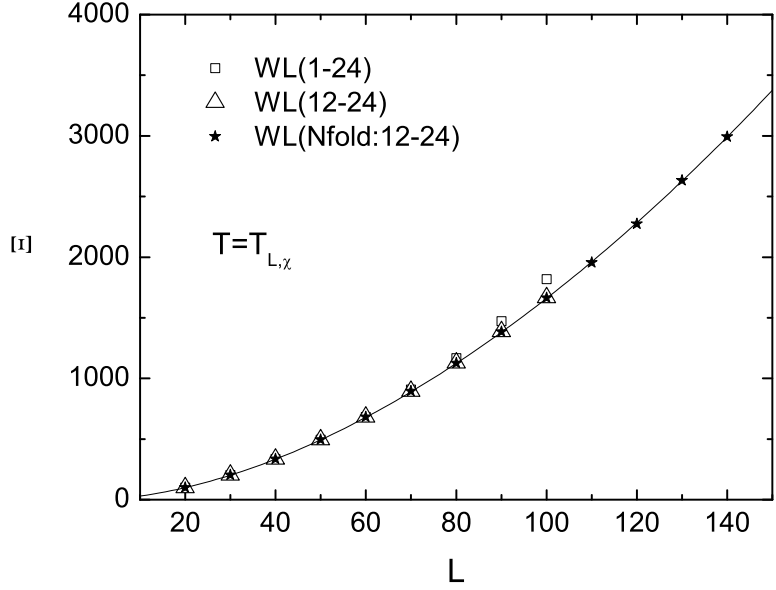}
\caption{\label{fig9} 2D Ising model: Finite-size behavior of the
extensions of critical minimum magnetic subspaces (CrMMS) obtained
from the WL schemes, at the susceptibility pseudocritical
temperatures, calculated with the help of the definition
(\ref{eq:19}) and using $r=10^{-6}$. The fitted line correspond to
the power law $0.525\cdot L^{1.75}$.}
\end{figure}

\begin{figure}
\includegraphics*[width=12 cm]{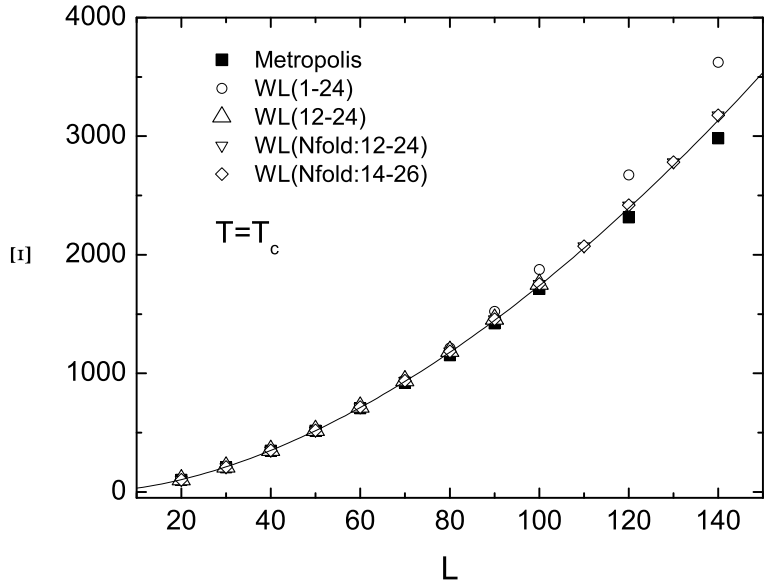}
\caption{\label{fig10} The same as in Fig.~\ref{fig8} at the exact
critical temperature $T_{c}$, including now for comparison the
CrMMS corresponding to the order parameter probability
distributions obtained by the Metropolis algorithm. The fitted
line, used as a guide to the eye, is the power law $0.55\cdot
L^{1.75}$, obtained by fixing the exponent to $1.75$ and fitting
the data $L=20-140$ of the WL(N-fold:12-24) scheme. Note the
decline of the Metropolis CrMMS.}
\end{figure}

Fig.~\ref{fig9} illustrates the scaling behavior of the critical
minimum magnetic subspaces (CrMMS) obtained using the magnetic
space restriction (\ref{eq:19}), defined above, and the accuracy
level $r=10^{-6}$ at the pseudocritical temperatures of the
susceptibilities. The behavior of the high-level WL recipes is
very good and provides quite accurate estimates for the critical
exponent $\gamma/\nu$. The line shown is the power law $0.525\cdot
L^{1.75}$ which is obtained by fixing the exponent to $1.75$ and
fitting the WL(N-fold:12-24) data. This is almost identical with
the power law obtained by using the exponent as a free parameter,
which yields $0.535\cdot L^{1.746}$. Note that the deviations of
the WL(1-24) scheme are not observable for small lattices.
However, with increasing lattice size ($L>60$) they become quite
apparent. Fig.~\ref{fig10} presents a similar illustration at the
exact critical temperature comparing now not only the WL recipes
but also the Metropolis method. The deviations from the expected
scaling law are now quite apparent not only for the ``bad''
WL(1-24) recipe but also for the Metropolis method.
Table~\ref{table1} gives estimates of the critical exponent
obtained from the schemes shown in Fig.~{\ref{fig10}} using only
the intermediate data $L=50-100$ in which these deviations are
still moderate. From this table the overestimation of the WL(1-24)
scheme but also the tendency of underestimation of the Metropolis
scheme is quite obvious. The Metropolis data used for $L=20-100$
were obtained from the CrMMS developed at the end of the $300$
time-steps described in Sec.~\ref{section3b}. It is of some
interest to list here some details of the end-points of the
dominant magnetization subspaces. Let us consider the case $L=100$
as an example. The broad energy range used in our runs for the WL
process was ($ie=850,ie=2150$), where the counting of energy
states is given by $ie=(E+2N)/4+1$. The dominant energy ranges are
(a) at the pseudocritical temperature of the specific heat
($ie=1061(3),ie=1947(3)$) and (b) at the pseudocritical
temperature of the susceptibility ($ie=1142(3),ie=1959(3)$). The
maximum value of the magnetization sampled in the process was
$|M|/N=0.912$ and the minimum value $|M|/N=0$. Defining the
dominant magnetization space with the help of Eq.~(\ref{eq:19}) at
the pseudocritical temperature of the susceptibility, we locate
this subspace as ($|M|/N=0,\;|M|/N=0.816$). This shows that the so
defined dominant magnetization subspace (CrMMS) is a subset of the
sampled values. Note that for all lattice sizes the left-end of
the dominant magnetization subspace is $|M|/N=0$.

\begin{figure}
\includegraphics*[width=11 cm]{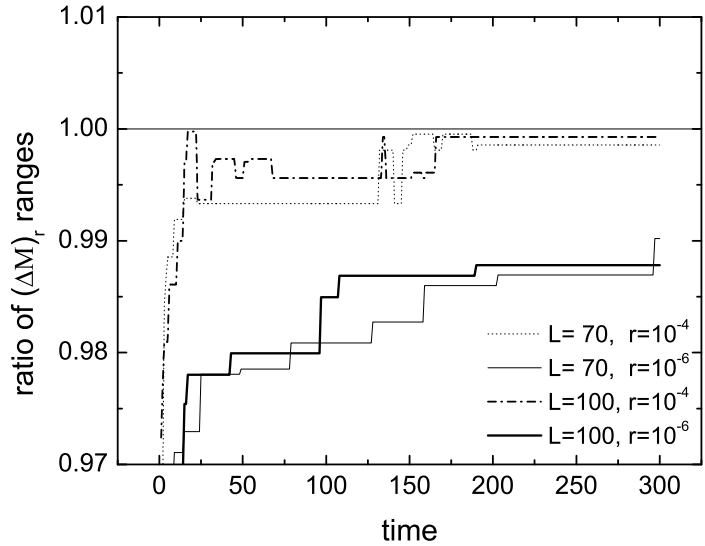}
\caption{\label{fig11} Illustration of the very slow equilibration
of the Metropolis algorithm in the tail regime. Time-development
of the extensions of CrMMS (corresponding to $r=10^{-4}$ and
$r=10^{-6}$) of the Metropolis algorithm. To observe the distance
from the true- or tail-equilibrium we have divided the
time-developing Metropolis extensions with the corresponding
extensions of the WL(N-fold:12-24) scheme so that true-equilibrium
occurs at the value $1$. It is obvious that for small $r$ and
large $L$ true-equilibrium of the Metropolis method is not
attained, even for very long runs.}
\end{figure}

\begin{figure}
\includegraphics*[width=12 cm]{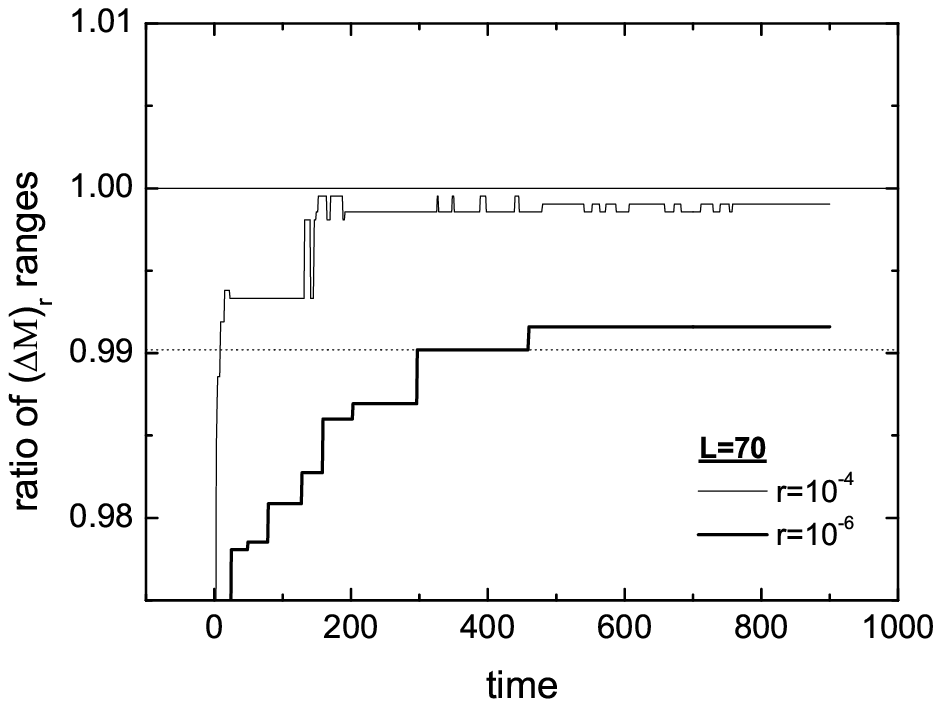}
\caption{\label{fig12} The same as in Fig.~\ref{fig11} for $L=70$
for a longer run. The dotted line shows the maximum value obtained
for $r=10^{-6}$ for the first $300$ time-steps (compare with
Fig.~\ref{fig11}).}
\end{figure}

Let us now present the Metropolis time-evolution of the dominant
$M$-subspaces since this development offers a didactic example of
the very slow tail-convergence of the traditional importance
sampling methods. Fig.~\ref{fig11} illustrates the slow
equilibration process of the Metropolis algorithm at the tails of
the order-parameter distribution. To draw this graph the
time-developing CrMMS, for the two $r$-levels ($r=10^{-4}$ and
$r=10^{-6}$) shown, was divided by the corresponding CrMMS
predicted by the high-level WL(N-fold:12-24) scheme which appear
to have very small errors for the lattice sizes shown. Thus,
considering these later CrMMS as exact the Metropolis relative
dominant $M$-subspaces should grow in time towards the value $1$.
For the value $r=10^{-4}$ this is almost achieved at the time
$t=150$ for both lattices shown, $L=70$ and $L=100$. However, for
$r=10^{-6}$, we observe a very slow relaxation process which
persists even if we increase the observation time. This is obvious
in Fig.~\ref{fig12} where, for the smaller lattice $L=70$, the
time duration of the process has been increased up to $t=900$.
Note that, if one was observing the equilibration process of the
algorithm with respect to the mean value of the order-parameter,
he would then have been convinced that equilibrium of this
quantity has been attained well before the time $t=150$.

This situation is due to the very slow equilibration at the tails
of the distribution and in particular at the large-$M$ tail. In
fact the time-expansion of the CrMMS in the Metropolis process is
due to the gradual movement of the right-end of the magnetization
range to larger values, since the left-end is $|M|/N=0$ from the
very first stages of the process. Returning to the large-$M$ range
is a rare event for the Metropolis algorithm and this makes this
traditional scheme inaccurate in the far tail-regime, but also
very inefficient for the study of the tails of the critical
distributions. The data points $L=120$ and $L=140$ for the
Metropolis algorithm included in Fig.~\ref{fig10} were obtained by
using runs with approximately equal time requirements (about $40$
hours in a Pentium IV 3GHz) with the WL(N-fold:14-26) scheme. For
the case $L=140$ this corresponds to only $30$ time-steps with the
developing relative CrMMS ($r=10^{-6}$) having hardy approached
the value of $0.97$ only. Even by using a much longer run (for
instance $100$ times longer) the Metropolis algorithm will not
give an adequate description of the far-tail regime.

\section{Concluding Remarks}
\label{section4}

Critical dominant energy and order-parameter subspaces can be
defined by various alternative restrictive schemes, as shown in
this work. In this way it is possible to optimize the Monte Carlo
schemes and study simultaneously all finite-size anomalies of
statistical models. The finite-size extensions of the dominant
energy and order-parameter subspaces scale with exponents
$\alpha/\nu$ and $\gamma/\nu$ respectively. Our experience with
this subject leads us to conclude that the extensions of these
dominant subspaces are more accurate than the estimates of the
corresponding thermodynamic variables (specific heat and
susceptibility), establishing the critical minimum subspace method
as a new alternative for the estimation of the associated critical
exponents from finite-size Monte Carlo data.

The presented clarification and generalization of the CRMES method
greatly speeds up the Monte Carlo simulations in many applications
of the methods determining the spectral densities in classical
statistical models. Furthermore, the presented one-run high-level
WL entropic sampling schemes provide efficient alternatives when
carried out in appropriately defined dominant subspaces. We expect
that for complex systems with long trapping times, these schemes
will appear to be much more advantageous in almost all respects.
This last expectation has been verified by our studies of the
random-field Ising model, which will be published shortly.
Moreover, these methods have general advantages, the most
important of these being: (a) the fact that one can improve the
$H(E,M)$ histograms by repeated application of the method (at the
same time we improve the accuracy of the DOS in the energy space),
and (b) the fact that their implementation in a sufficiently broad
energy subspace provides data for calculating all finite-size
properties of the statistical system, which are relevant for the
prediction of the asymptotic critical behavior. Overall, we
envisage that our study can be further utilized in many ways for
the investigation of the critical behavior of statistical models
in future researches.

\begin{acknowledgments}
The authors would like to thank a referee for stimulating
suggestions on the Wang-Landau method. This research was supported
by the Special Account for Research Grants of the University of
Athens under Grant No. $70/4/4071$.
\end{acknowledgments}

{}
\clearpage


\begin{thebibliography}{}

\bibitem{metropolis53} N. Metropolis, A. W. Rosenbluth, M. N. Rosenbluth, A. H.
Teller,
and E. Teller, J. Chem. Phys. {\bf 21}, 1087 (1953).

\bibitem{bortz75} A. B. Bortz, M. H. Kalos, and J. L. Lebowitz, J. Comput. Phys. {\bf 17},
10 (1975).

\bibitem{binder77} K. Binder, Rep. Prog. Phys. {\bf 60}, 487 (1977).

\bibitem{newman99} M. E. J. Newman and G. T. Barkema, \textit{Monte Carlo Methods in
Statistical Physics} (Clarendon Press, Oxford, 1999).

\bibitem{landau00} D. P. Landau and K. Binder, \textit{A Guide to Monte Carlo Simulations in
Statistical Physics} (Cambridge University Press, 2000).

\bibitem{dayal04} P. Dayal, S. Trebst, S. Wessel, D. W\"{u}rtz, M.
Troyer, S. Sabhapandit, and S. N. Coppersmith, Phys. Rev. Lett.
{\bf 92}, 097201 (2004).

\bibitem{lee93} J.Lee, Phys. Rev. Lett. {\bf 71}, 211 (1993).

\bibitem{berg92} B. A. Berg and T. Neuhaus, Phys. Rev. Lett. {\bf 68}, 9 (1992).

\bibitem{lima00} A. R. Lima, P. M. C. de Oliveira, and T. J. P.
Penna, J. Stat. Phys. {\bf 99}, 691 (2000).

\bibitem{kastner00} M. Kastner, M. Promberger, and J. D. Munoz,
Phys. Rev. E {\bf 62}, 7422 (2000).

\bibitem{wang02} J. -S. Wang and R. H. Swendsen, J. Stat. Phys. {\bf
106}, 245 (2002); J. -S. Wang, T. K. Tay, and R. H. Swendsen,
Phys. Rev. Lett. {\bf 82}, 476 (1999).

\bibitem{wang01} F. Wang and D. P. Landau, Phys. Rev. Lett. {\bf 86}, 2050 (2001); Phys. Rev. E {\bf 64}, 056101 (2001).

\bibitem{landau04} D. P. Landau and F. Wang, Braz. J. Phys. {\bf 34}, 354 (2004).

\bibitem{malakis04} A. Malakis, A. S. Peratzakis, and N. G. Fytas, Phys. Rev. E {\bf 70}, 066128 (2004).

\bibitem{martinos05} S. S. Martinos, A. Malakis, and I. A. Hadjiagapiou, Physica A {\bf 352}, 447 (2005).

\bibitem{challa86} M. S. S. Challa, D. P. Landau, and K. Binder, Phys. Rev. B {\bf 34}, 1841 (1986).

\bibitem{binder84} K. Binder and D. P. Landau, Phys. Rev. B {\bf 30}, 1477 (1984).

\bibitem{ferdinand69} A. E. Ferdinand and M. E. Fisher, Phys. Rev. {\bf 185}, 832 (1969).

\bibitem{malakis04b} A. Malakis, S. S. Martinos, I. A. Hadjiagapiou, and A. S. Peratzakis, Int.
J. Mod. Phys. C {\bf 15}, 729 (2004).

\bibitem{zhou05} C. Zhou and R. N. Bhatt, Phys. Rev. E {\bf 72},
025701 (2005).

\bibitem{lee} H. K. Lee, Y. Okabe, and D. P. Landau,
cond-mat/0506555.

\bibitem{oliveira00} P. M. C. de Oliveira, Braz. J. Phys. {\bf 30}, 195 (2000).

\bibitem{schulz01} B. J. Schulz, K. Binder, and M. Muller, Int. J. Mod.
Phys. C {\bf 13}, 477 (2001).

\bibitem{pomeau84} Y. Pomeau, J. Phys. A {\bf 17}, 415 (1984); G.
Vichniac, Physica D {\bf 10}, 96 (1984).

\bibitem{herrmann86} H. J. Herrmann, J. Stat. Phys. {\bf 45}, 145
(1986); J. G. Zabolitzky and H. J. Herrmann, J. Comput. Phys. {\bf
76}, 426 (1988).

\bibitem{creutz83} M. Creutz, Phys. Rev. Lett. {\bf 50}, 1411
(1983).

\bibitem{berg93} B. Berg, Nature {\bf 361}, 708 (1993).

\bibitem{jswang99} J. -S. Wang, Eur. Phys. J {\bf 8}, 287 (1999);
Physica A {\bf 281}, 147 (2000).

\bibitem{trebst04} S. Trebst, D. A. Huse, and M. Troyer, Phys. Rev.
E {\bf 70}, 046701 (2004).

\end{thebibliography}
\end{document}